\documentclass[leqno,a4paper,12pt]{article}
\usepackage[centertags]{amsmath}
\usepackage{amsfonts}
\usepackage{amssymb}
\usepackage{multirow}
\usepackage{amsthm}
\usepackage{color}
\usepackage{nicefrac}
\usepackage{pifont}
\usepackage[ansinew]{inputenc}
\usepackage{graphicx}

\theoremstyle{plain}

\begin{document}

\title{State model for partly undetected non-communicable diseases (NCDs)}
\author{Ralph Brinks\\German Diabetes Center}
\maketitle

\begin{abstract}
This article proposes an age-structured compartment model for 
irreversible diseases with a pre-clinical state of undiagnosed cases that
precedes the diagnosis. The model is able to cope with mortality rates 
differing between the pre-clinical and the clinical state
(differential mortality). Applicability is tested in a hypothetical
disease with realistic incidence and mortality rates.
\end{abstract}

\section{Introduction}
Many non-communicable diseases (NCDs) progress unobserved before clinical
symptoms occur. Examples are certain types of cancer \cite{Lao13},
diabetes \cite{Bea14} and dementia \cite{Hod11}. Mostly, reasons for the undetected
progression may be the lack of early symptoms. Other reasons are missing
awareness of patients and physicians, lack of practical diagnostic
tests or incoherent definitions for a diagnosis of the disease. In
some cases, the progressing condition may be completely unknown.

\bigskip

The next section introduces a compartment model with a pre-clinical stage
preceding the clinical stage. As in the field of infectious disease
epidemiology we describe the disease dynamics of a population by differential 
equations involving the transition rates between the compartments \cite{Vyn10}.
The model described here is able to cope with secular trends, i.e. involves
calendar time $t$, and the different ages $a$ of the subjects in the population. Sometimes
these models are called \emph{age-structured}. 

Although considering a pre-clinical state preceding a diagnosis is
at least going back to 1969 \cite{Zel69} and a considerable amount
of work has been devoted to compartment models since then, to our knowledge a
description using differential equations in calendar time $t$ and age $a$
is new. Furthermore, 
other models distinguish between disease-specific mortalities 
and other causes of death (e.g. \cite{Tol78}). However, we consider
this approach critical, because there might be cases where the cause of
death is not clearly attributable. Is a death by an infection (say pneumonia)
due to the disease (e.g. the immunosuppressive treatment in cancer) or is it
independent from the disease? In practical cases this is difficult to judge.

\bigskip

In the third section the compartment model is used in an example
of an partly unobserved disease. We mimic the situation that a hypothetical 
population suffers from a fictional NCD that is detected at a certain point
in time. From that time on, the medical community is aware, starts
to diagnose and treat the newly discovered condition.

\section{Diagnosis Model}
\begin{figure}[ht]
  \centering
  \includegraphics[keepaspectratio,width=0.9\textwidth]{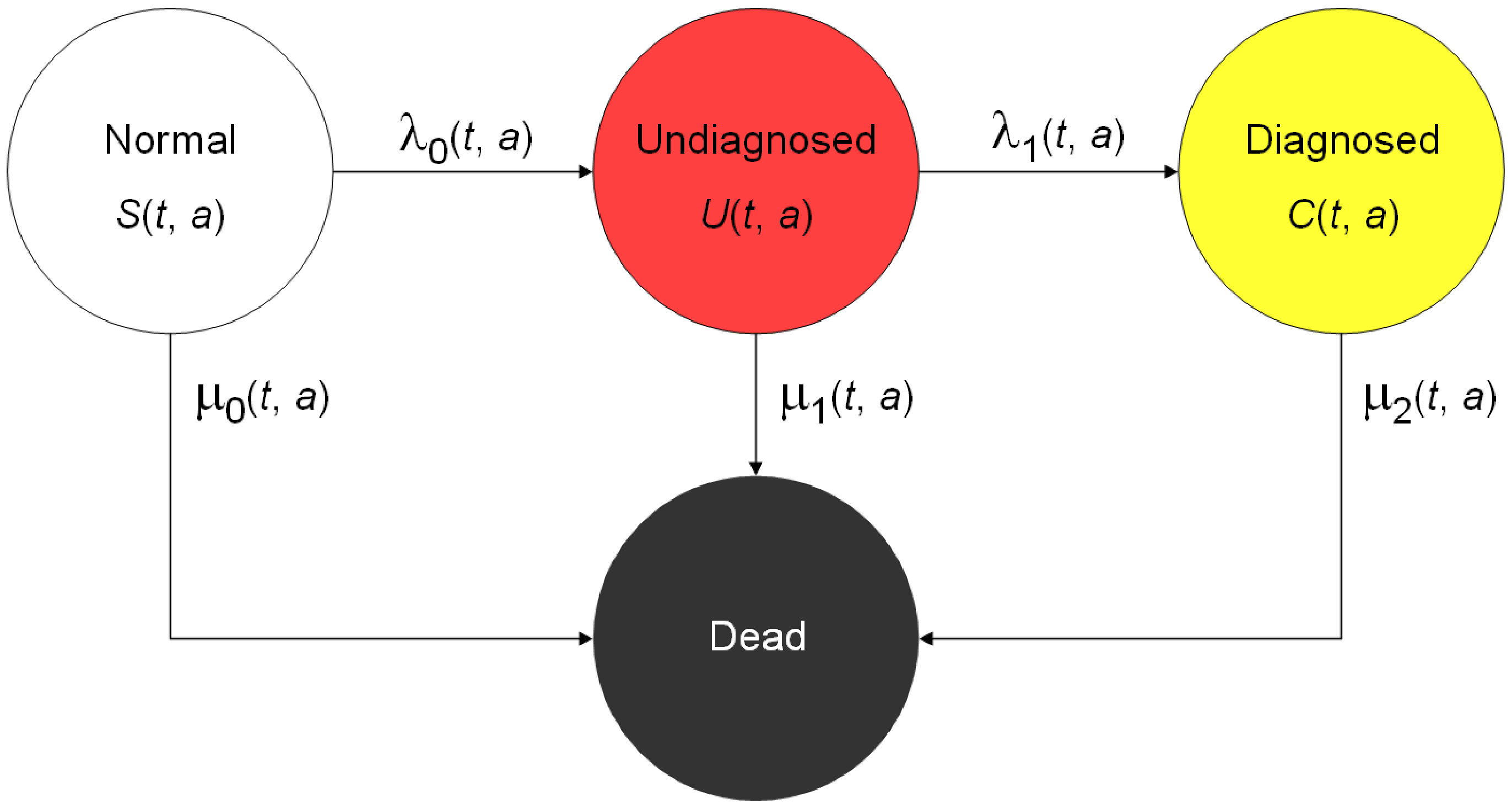}
\caption{Chronic disease model with four states and the
corresponding transition rates. People in the state \emph{Normal}
are healthy with respect to the disease under consideration. After
onset of the disease, they change to state \emph{Undiagnosed} and 
later to the state \emph{D}. The absorbing state \emph{Dead} can be
reached from all other states. The numbers of persons in the states and
the transition rates depend on calendar time $t$ and age $a.$}
\label{fig:CompModel}
\end{figure}

In modelling chronic (irreversible) diseases, often the
three-state model (compartment model) in Figure
\ref{fig:CompModel} is used. The numbers of persons in the states
\emph{Normal, Undiagnosed} and \emph{Diagnosed} are denoted by $S, U,$ and $C$. The
transition intensities (synonym: rates) between the states are:
the incidence rates $\lambda_0, \lambda_1$ and the mortality rates $\mu_0, \mu_1$ and 
$\mu_2.$ These rates generally depend on the calendar time $t$ and the age $a$.

Although the inclusion of the disease duration $d$ is also
possible \cite{Bri13}, hereinafter it is assumed that $m_1$ does not depend on
$d.$ Analogously to \cite{Bri12b}, we look for the numbers $S(t, a), U(t, a)$ and $C(t,
a)$ of healthy, undiagnosed and diagnosed persons in terms of differential
equations, which can be derived from the disease model in Figure
\ref{fig:CompModel}. For the healthy persons we get the following
initial value problem of Cauchy type:
\begin{align}
(\partial_t + \partial_a) \, S(t, a) & = - \bigl ( \mu_0(t, a) +
\lambda_0(t, a) \bigr ) \, S(t, a) \label{e:PDE_S_ta} \\
S(t, 0) & = S_0(t). \nonumber
\end{align}

Here $S_0(t)$ is the number of (healthy)
newborns\footnote{Here we just consider diseases contracted
after birth.} at calendar time $t.$ The notation $\partial_x$
denotes the partial derivative with respect to $x, ~x \in \{t,
a\}$.

\bigskip

The numbers $U$ and $C$ of diseased persons without and with diagnosis 
are described similarly:
\begin{align}
(\partial_t + \partial_a) \, U(t, a) & = - \bigl ( \mu_1(t, a) + \lambda_1(t, a) \bigr ) \, U(t, a) +
\lambda_0(t, a)\, S(t, a) \label{e:PDE_U_ta} \\
U(t, 0) & = 0. \nonumber \\
(\partial_t + \partial_a) \, C(t, a) & = -\mu_2(t, a) \, C(t, a) +
\lambda_1(t, a)\, U(t, a) \label{e:PDE_C_ta} \\
C(t, 0) & = 0. \nonumber
\end{align}

After defining $N(t, a) = S(t, a) + U(t, a) + C(t, a)$ and 
\begin{align*}
p_0(t, a) &= \nicefrac{S(t, a)}{N(t, a)} \\
p_1(t, a) &= \nicefrac{U(t, a)}{N(t, a)} \\
p_2(t, a) &= \nicefrac{C(t, a)}{N(t, a)},
\end{align*}
the overall mortality (general mortality) $\mu$ in the population may be written as
\begin{equation*}
\mu = p_0 \, \mu_0 + p_1 \, \mu_1 + p_2 \, \mu_2.
\end{equation*}

By using $p_0 + p_1 + p_2 = 1,$ the partial differential equations 
\eqref{e:PDE_U_ta} and \eqref{e:PDE_C_ta} read as

\begin{align}
(\partial_t + \partial_a) p_1 &= - \bigl ( \lambda_0 + \lambda_1 + \mu_1 - \mu \bigr ) \, p_1 - \lambda_0 \, p_2 + \lambda_0 \label{e:pde1}\\
(\partial_t + \partial_a) p_2 &= \lambda_1 \, p_1 - \bigl ( \mu_2 - \mu \bigr ) \, p_2 \label{e:pde2}.
\end{align}

Together with the initial conditions $p_1(t , 0) = p_2(t , 0) = 0$ for all $t,$ the
system \eqref{e:pde1} - \eqref{e:pde2} completely describes the dynamics of the disease 
in the considered population. The values of $p_0$ are obtained by using $p_0 = 1- p_1 - p_2.$
Note that the system \eqref{e:pde1} - \eqref{e:pde2} does not explicitly depend on the mortality
of the healthy subjects $\mu_0,$ which is typically unknown. The remaining rates 
($\lambda_0, \lambda_1, \mu_1, \mu_2$) are either
accessible by (specially designed) epidemiological studies 
or by official vital statistics ($\mu$).


\section{Simulation}
We use system \eqref{e:pde1} - \eqref{e:pde2} to describe a hypothetical
irreversible disease, which is undiagnosed until a specific point in time $t^\star.$
At $t^\star$ the disease is detected and diagnosed henceforth. As a consequence,
after $t^\star$ the prevalence $p_1$ of undetected cases decreases whereas the prevalence $p_2$ of
detected cases increases. The general mortality $\mu$ is chosen
as the (approximated) general mortality of the German male population 
in from 1900 ($t = 0$) to 2010 ($t = 110:$) 

$$\mu (t, a) = \exp \bigl ( \beta_0(t) + \beta_1(t) \, a \bigr ),$$
with $\beta_0(t) = - 7.078 - 0.02592 \, t$ and 
$\beta_1(t) = 0.06401 + 2.455 \, 10^{-4} \, t.$ For simplicity, the mortality rates
$\mu_\ell, ~\ell = 1,2,$ are assumed to be proportional to $\mu:$ $\mu_1 = 3.5 \, \mu$ 
and $\mu_2 = 2.5 \, \mu.$ The factor for $\mu_1$ is chosen to be larger than the
one for $\mu_2,$ because in contrast to the the persons in the \emph{detected} state
the persons in the \emph{undetected} state cannot be treated for the disease.

\medskip

The rates $\lambda_\ell, \ell = 0,1,$
are modified incidence rates of dementia in German males \cite{Zie09}.
The rate $\lambda_0$ is the 1.5-fold rate of the values in \cite{Zie09}, which mimics
one undetected case per two detected cases for $t \ge 75,$ see Table \ref{tab}. 
For year $t = 75$ the rates $\lambda_1$ are also shown in Table \ref{tab}. There is
a secular trend in $\lambda_1$ mimicking the increasing awareness for the disease.
In the simulation, $\lambda_1$ increases by 1 \% per year for all ages $a.$

\begin{table}[ht]
  \centering
\begin{tabular}{c|cc}
Age        & Incidence $\lambda_0$  & Incidence $\lambda_1$ in the year 75\\
(years)    & (per 100 person-years) &  (per 100 person-years) \\\hline
$\le 62.5$  &  0  & 0   \\
  67.5     &  0.3 & 3.3  \\
  72.5     &  0.7 & 7.8  \\
  77.5     &  1.7 & 20   \\
  82.5     &  3.0 & 33   \\
  87.5     &  5.2 & 58   \\
  92.5     &  7.6 & 510  \\
  97.5     &  9.9 & 1340 \\
$\ge 100$  & 11.2 & 4110 \\
\end{tabular}
\caption{Age-specific incidence rates $\lambda_0$ and $\lambda_1.$ For the $t>75$
the rate $\lambda_1$ increases by 1\% annually for all ages.}\label{tab}
\end{table}

\begin{figure}[ht!]
  \centering
  \includegraphics[width=.9\textwidth,keepaspectratio]{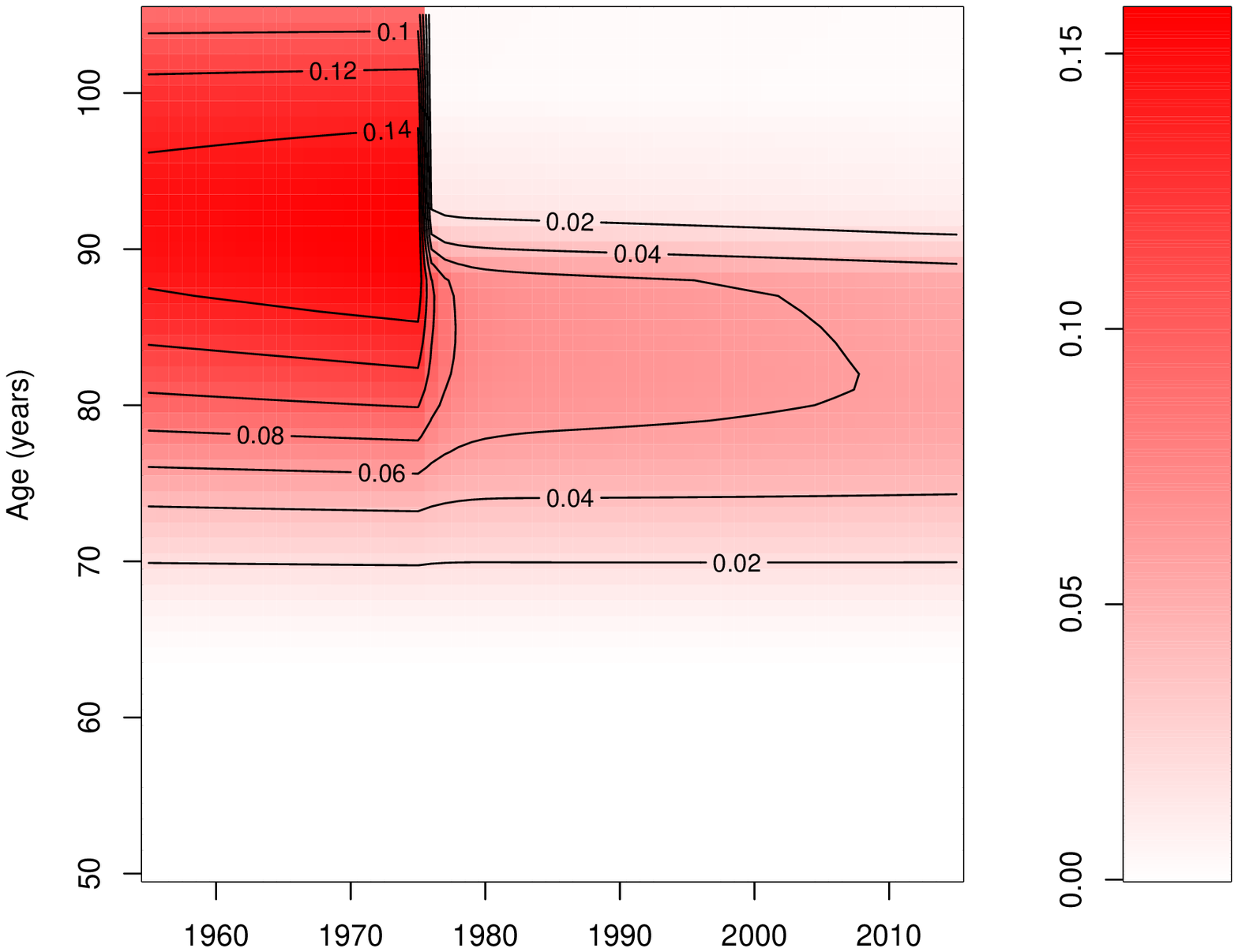}\\
  \caption{Prevalence of the undiagnosed disease ($p_1$) over year and age (left). The
  colour corresponds to value of the prevalence (coding scheme on the right hand side).}\label{fig:undiag}
\end{figure}

\begin{figure}[ht!]
  \centering
  \includegraphics[width=.9\textwidth,keepaspectratio]{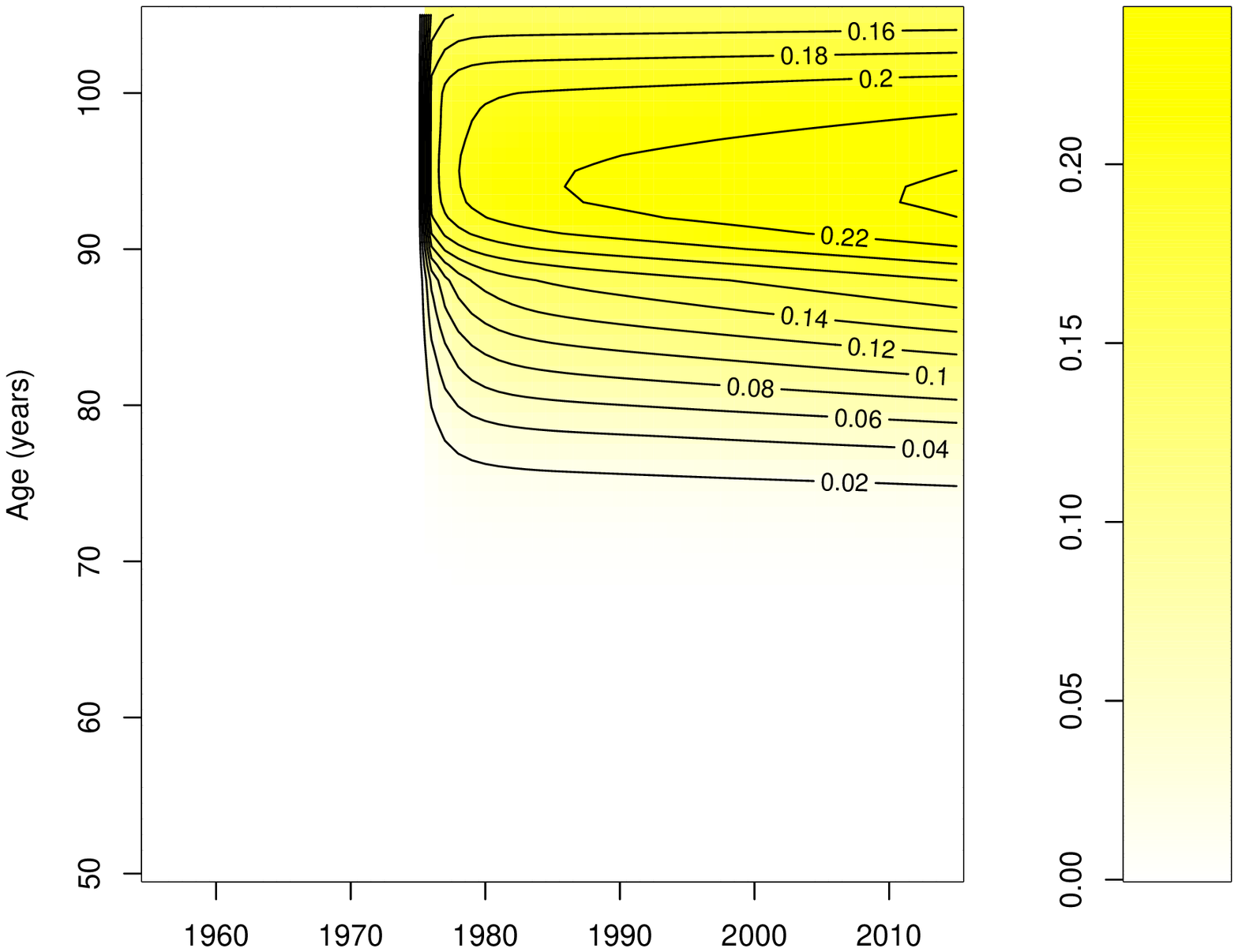}\\
  \caption{Prevalence of the diagnosed disease ($p_2$) over year and age (left). The
  colour corresponds to the value of prevalence (coding scheme on the right hand side).}\label{fig:diag}
\end{figure}

If we solve the system \eqref{e:pde1} - \eqref{e:pde2} by the methods of
characteristics, we obtain the prevalences of the
undiagnosed and diagnosed disease as shown in Figures \ref{fig:undiag} and
\ref{fig:diag}, respectively. The qualitative change after 1975 ($t = 75$) 
in both prevalences $p_1$ and $p_2$ is clearly visible in the upper right corner
of the figures.

\begin{figure}[ht!]
  \centering
  \includegraphics[width=.95\textwidth,keepaspectratio]{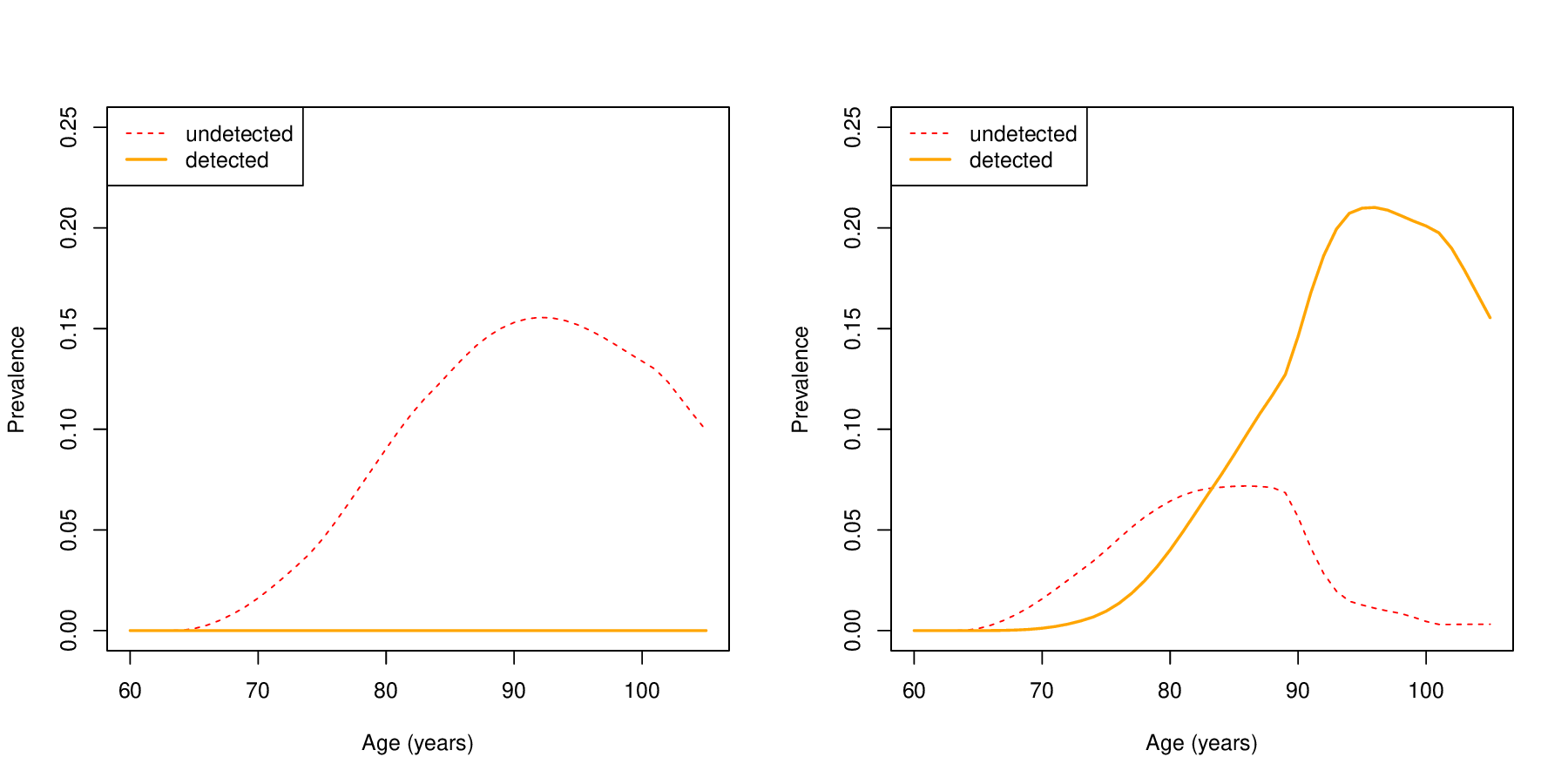}\\
  \caption{Age-specific prevalence of the undiagnosed (red, dashed lines) and 
  the diagnosed disease (orange, solid lines) in 1970 ($t = 70$, left) and
  in 1980 ($t=80$, right).}\label{fig:cross}
\end{figure}

For better comparison, the age-specific prevalences in 1970 ($t=70$) and
1980 ($t=80$) are additionally shown in Figure \ref{fig:cross}. In 1970, there
are no diagnosed cases (the hypothetical disease is not detected yet). 
The prevalence of the undiagnosed cases ($p_1$) is peaking at
about 16\% at the age of 91 years. Ten years later, the disease has been detected
and the medical community is making diagnoses. Hence, the prevalence of the
undiagnosed disease has tremendously decreased -- to less than 8\%. Especially in the
higher age groups ($\ge 95$) the physicians are aware and detect a high proportion
of cases. Thus, the prevalence of diagnosed cases ($p_2$) has increased a lot.

\begin{figure}[ht!]
  \centering
  \includegraphics[width=.90\textwidth,keepaspectratio]{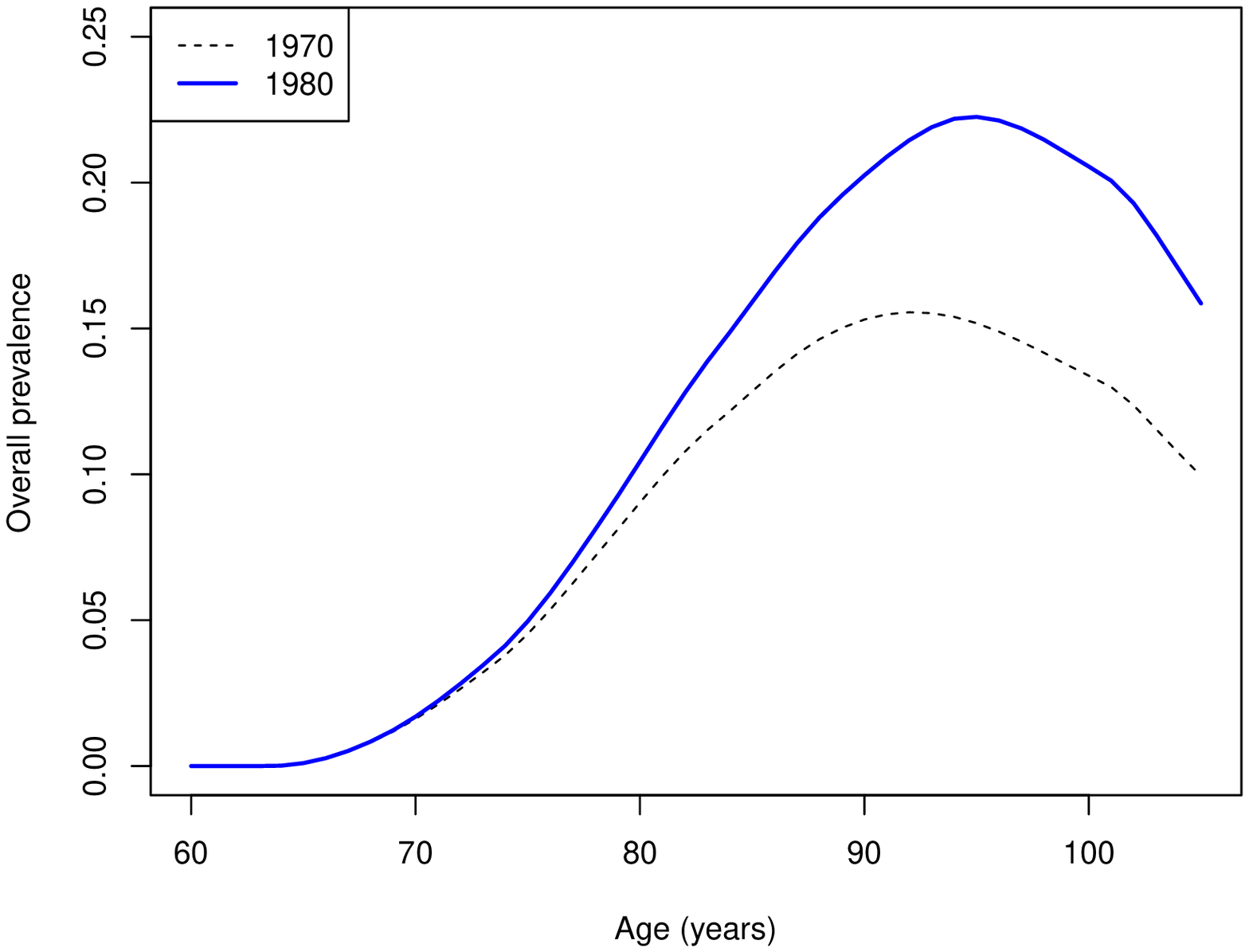}\\
  \caption{Overall age-specific prevalence ($p_1 + p_2$) in 1970 (black dashed line) and
  in 1980 (blue solid line).}\label{fig:overall}
\end{figure}

It is
amazing, how much the overall prevalence ($p_1 + p_2$) in 1970 differs from the one
in 1980 (cf. Figure \ref{fig:overall}). This is an effect of the lowered mortality
for those diseased persons who have been detected (and are treated since
then). Since the mortality $\mu_2$
is much lower than $\mu_1,$ the overall survival of the diseased persons is 
improved after 1975 and the overall prevalence increases.

\clearpage


\section{Discussion}
In this work, a novel approach for analysing epidemiological
measures of a chronic disease is proposed. With a view
to the chronic (incurable) disease, a pre-clinical state is considered
in which the disease is at least partly undiagnosed. The situation is
described in an age-structured compartment model unsing a set of partial 
differential equations.

There are various chronic diseases that have an pre-clinical state 
preceding a diagnosis.
Examples were given in the introductory section of this article. Other diseases
with an asymptomatic pre-clinical state are chronic kidney disease 
(CKD), hypertension and arteriosclerosis. 

So far, we just considered 
non-communicable diseases. 
However some incurable infectious diseases, such as HIV or hepatitis C, 
also have an asymptomatic pre-clinical phase. Thus, the compartment
model of Figure \ref{fig:CompModel} as well may be useful in these cases.

In an example we have demonstrated the applicability of the modelling framework
for a hypothetical chronic disease that has been detected at a specific point
in time and has been diagnosed and treated since then. Examples analogue to
the one shown may give insight about the ratio between undiagnosed and diagnosed
cases in a chronic disease.

A final remark about the system \eqref{e:pde1} - \eqref{e:pde2} and the example:
Since the transition rates for the compartment model are assumed to be known, 
we speak of a \emph{forward problem} \cite{Bri12a}. There is an associated 
\emph{inverse problem}, which might be interesting.


{}

\bigskip

\emph{Contact:} \\
Ralph Brinks \\
German Diabetes Center \\
Auf'm Hennekamp 65 \\
D- 40225 Duesseldorf\\
\verb"ralph.brinks@ddz.uni-duesseldorf.de"
\end{document}